\documentclass[a4paper,11pt]{article}
\pdfoutput=1 

\usepackage{jheppub} 
\graphicspath{{Figures/}}

  \newcommand{\be}{\begin{equation}}
  \newcommand{\bea}{\begin{eqnarray}}
  \newcommand{\eea}{\end{eqnarray}}
  \newcommand{\beq}{\begin{equation}}
  \newcommand{\ee}{\end{equation}}
  \newcommand{\eeq}{\end{equation}}

\title{Free energy landscape and kinetics of phase transition in two coupled SYK models}
\author[a,b]{Ran Li,}
\author[b,c,*]{Jin Wang \note[*]{Corresponding author}}
\affiliation[a]{School of Physics, Henan Normal University, Xinxiang 453007, China}
\affiliation[b]{Department of Chemistry, Stony Brook University, Stony Brook, NY 11794, USA}
\affiliation[c]{Department of Physics and Astronomy, Stony Brook University, Stony Brook, NY 11794, USA,}

\emailAdd{liran@htu.edu.cn}
\emailAdd{jin.wang.1@stonybrook.edu}

\abstract{We propose that the thermodynamics and the kinetics of the phase transition between wormhole and two black hole described by the two coupled SYK model can be investigated in terms of the stochastic dynamics on the underlying free energy landscape. We assume that the phase transition is a stochastic process under the thermal fluctuations. By quantifying the underlying free energy landscape, we study the phase diagram, the kinetic time and its fluctuations in details, which reveal the underlying thermodynamics and kinetics. It is shown that the first order phase transition between wormhole and two black hole described by two coupled SYK model is analogous to the Van der Waals phase transition. Therefore, the emergence of wormhole and two black hole phases, the phase transition and associated kinetics can be quantitatively addressed in our free energy landscape and kinetic framework through the dependence on the barrier height and the temperature. }

\begin{document} 

\maketitle

\section{Introduction}

 Originally, wormholes are the imaginary objects that are used for time travels or fast-than-light journeys in science fictions. However, it is shown that wormholes \cite{Einstein:1935tc,Fuller:1962zza,Morris:1988tu,Visser:1995cc} as the special spacetime structures play more and more important role in understanding the physics of our universe. Recent examples include: the Maldacena and Susskind's proposal of "ER=EPR" conjecture \cite{Maldacena:2013xja}, the traversable wormholes from the double trace coupling between the boundaries of AdS black hole \cite{Gao:2016bin}, Hayden-Preskill protocol realization \cite{Hayden:2007cs} or quantum teleportation towards traversable wormholes \cite{Maldacena:2017axo}, replica wormholes used to derive the island rule for the entanglement entropy of Hawking radiation \cite{Almheiri:2020cfm,Penington:2019kki,Almheiri:2019qdq}, and humanly traversable wormhole solutions in general relativity \cite{Blazquez-Salcedo:2020czn,Maldacena:2020sxe}. Wormholes have attracted significant research attentions in recent years.  
 
 Traversable wormholes constructed by turning on the interaction that couples the two boundaries of an eternal AdS black hole \cite{Gao:2016bin} is an important example that may have an interesting interpretation in the context of ER=EPR conjecture. In particular, Maldacena and Qi \cite{Maldacena:2018lmt} have constructed the nearly-$AdS_2$ solution \cite{Almheiri:2014cka,Jensen:2016pah,Maldacena:2016upp,Engelsoy:2016xyb} describing an eternal traversable wormhole by introducing the boundary interaction in Jackiw-Teitelboim (JT) gravity \cite{Teitelboim:1983ux,Jackiw:1984je}. It is also shown that the system undergoes a first order transition at finite temperature from the wormhole phase to a two black holes phase, which displays the usual Hawking-Page phase transition \cite{Hawking:1982dh} between the thermal AdS phase at low temperature and the black hole phase at high temperature. They further argued that, in the dual two Sachdev-Ye-Kitaev (SYK) systems \cite{Sachdev:1992fk,Kitaev,Kitaev:2017awl,Maldacena:2016hyu} coupled by a relevant interaction, there is a similar phase diagram. The following analytical and numerical calculations confirm this argument \cite{Horowitz:2019hgb,Qi:2020ian,Zhang:2020szi,Sorokhaibam:2019qho,Garcia-Garcia:2020ttf,Maldacena:2019ufo}.

 In the present work, for the two coupled SYK model at the large $N$ limit, according to the thermodynamics of the system, we define the off-shell free energy and quantify the corresponding free energy landscape. In the framework of the free energy landscape \cite{FSW,FW,NG,JW}, we analyze the phase diagram between the wormhole and two black hole described by the two coupled SYK model and obtain the conclusion that the phase transition in this model is analogous to the Van der Waals type phase transition \cite{DCJ,Kubiznak:2012wp,Li:2020nsy}, not exactly the Hawking-Page phase transition \cite{Li:2020khm}. Furthermore, we propose the kinetics of the phase transition between the wormhole and two black hole can be investigated by using the stochastic dynamics on the free energy landscape. We assume the phase transition between the wormhole and two black hole is a stochastic process under the thermal fluctuations and study the probabilistic Fokker-Planck equation which describes the kinetics of the phase transition \cite{Li:2020nsy,Li:2020khm,Li:2021vdp,Wei:2020rcd,Li:2020spm,Wei:2021bwy,Cai:2021sag,Lan:2021crt,Li:2021zep,Yang:2021nwd,Mo:2021jff,Kumara:2021hlt}. By calculating the kinetics time and its fluctuation, we reveal the underlying thermodynamics and kinetics of the phase transition between the wormhole and two black hole. It is shown that the underlying reason of the kinetic behavior is determined by the barrier heights and the temperature on the free energy landscape.  
 
 We emphasize that the first order phase transition between the wormhole and two black hole in two coupled SYK model is not Hawking-Page type transition \cite{Li:2020khm}. Our argument is based on the following two points. The shape of the free energy landscape is double well, which is different from the shape of the Hawking-Page transition \cite{Li:2020khm}. However, one of the double well here locates at the position where the order parameter is zero. This is very similar to the free energy landscape of the Hawking-Page transition where one of the well is at zero. In this sense, both hawking-Page type landscape and double wells describe first order phase transition. The second point is the existence of the critical point as in the gas-liquid phase transition in van der waals fluid \cite{DCJ,Kubiznak:2012wp,Li:2020nsy}. Therefore, we conclude the first order phase transition in two coupled SYK model is analogous to the Van der Waals phase transition rather than Hawking-Page type.
 
 The rest of the paper is arranged as follows. In section \ref{model}, we review the basic fact about the two coupled SYK model and its thermodynamics at the large $N$ limit. In section \ref{landscape}, we define the off-shell free energy and quantify the free energy landscape. Using the free energy landscape topography, we analyse the phase diagram between the wormhole and two black hole of the system. In section \ref{kinetics}, using the Fokker-Planck equation, we calculate the kinetic time and its fluctuation as the functions for the phase transition of the temperature and study the underlying reason. The summary and discussion are presented in the last section.

\section{Two coupled SYK model}\label{model}

The SYK model has a Hilbert space generated by $N$ Majorana fermions $\psi^i$ with a Hamiltonian of the form \cite{Sachdev:1992fk,Kitaev}
\bea \label{SingleSYK}
H &=& (i) ^{ q/2} \sum_{1 \leq j_1 \leq j_2  \cdots \leq  j_q } J_{j_1 j_2 \cdots j_q} \psi^{j_1} \psi^{j_2} \cdots \psi^{ j_q }  ~,
\cr
 & ~& \langle J_{j_1 \cdots j_q }^2 \rangle =
 %  { J^2 (q-1)! \over N^{q-1} } =
  { 2^{ q -1}  {\cal J } ^2 (q-1)! \over q N^{q-1} } ~~~ ({ \rm no ~sum } )
\eea
where the couplings are drawn from a random gaussian distribution with the mean indicated above. The factor of $i$ is necessary to maintain a hermitian Hamiltonian when $q/2$ is odd.

The Hamiltonian of the two coupled SYK model is given by \cite{Maldacena:2018lmt} 
\be \label{Hint}
H_{\rm total} = H_{\rm L, SYK} + H_{\rm R,SYK} + H_{\rm  int} ~,~~~~~~H_{\rm int} = i \mu \sum_j    \psi_L^j \psi_R^j
\ee
where $H_{\rm L, SYK}$ and $H_{\rm R,SYK}$ are the Hamiltonians of the two decoupled SYK models and $H_{\rm int}$ denotes the interaction of the two copies of the SYK model with $\mu$ being the coupling constant. We denote the fermions of the two copies the SYK model as $\psi^i_L$ and $\psi^i_R$. For the decoupled models, $H_{\rm L, SYK}$ and $H_{\rm R,SYK}$ are described by the same couplings, up to a sign for odd $q/2$,
$J_{j_1 \cdots j_q} ^L = (-1)^{q/2} J^R_{j_1 \cdots j_q}$.

 We consider the large $N$ case. The effective action can be written as \cite{Maldacena:2018lmt}
 \begin{eqnarray} \label{EfAct}
 -S_E/N&=&\log{\rm Pf}\left(\partial_\tau\delta_{ab}-\Sigma_{ab}\right)-\frac12\int d\tau_1d\tau_2\sum_{a,b}\left[\Sigma_{ab}(\tau_1,\tau_2)G_{ab}(\tau_1,\tau_2)-s_{ab}\frac{ {\cal J}^2}{2 q^2}[ 2 G_{ab}(\tau_1,\tau_2)]^q\right] + \nonumber\\
 & &+ { i \mu \over 2}  \int d\tau_1 \left[ - G_{LR}(\tau_1,\tau_1) + G_{RL}(\tau_1,\tau_1) \right]
 \end{eqnarray}
 Here $a,b=L,R$ denotes the two sides. Note that the functions obey the antisymmetry condition $G_{ab}(\tau_1,\tau_2) = - G_{ba}(\tau_2,\tau_1) $.
 Here $s_{ab}$ denotes a sign, $s_{LL} = s_{RR} =1$, $s_{LR} = s_{RL} =(-1)^{q/2} $, which arises because for odd $q/2$ the signs of the couplings of the left Hamiltonian are the opposite compared to those of the right Hamiltonian (with the same absolute value). The Green's functions satisfy the Schwinger-Dyson equations \cite{Maldacena:2018lmt}, which are not listed for simplicity. 
 
 In the large $q$ limit, the effective action is amenable to an analytical calculation \cite{Maldacena:2018lmt}. When the inverse temperature $\beta$ is of order $q \log q$, the logarithm of the partition function $ \log Z/N$ is given by \cite{Maldacena:2018lmt}
 \be \label{partfunc}
 {\log Z \over N} =   {   \tanh \tilde \gamma  \log(q/\sigma) \over q  } \left[  { q \over 2 } - 1 + { 1 \over \tanh \gamma \tanh \tilde \gamma } + \log { \sinh \gamma \over \cosh \tilde \gamma } + { \sigma \over \tanh \tilde \gamma } \right] + { \sigma \over q },
\ee 
where the parameters are defined as 
\bea \label{BCTh}
&&\sigma = q e^{ - \beta \nu }~,~~~~
\nu = {2 \alpha \over q }={ \mu \over \tanh \tilde \gamma} ~,~~~~
 \alpha = {\cal J} \sinh \gamma ~,~~~~~
 \nonumber\\
 &&
 \tilde \gamma = \gamma + \sigma ~,~~~~ \hat \mu = \mu q = 2 \alpha \tanh \tilde \gamma  ~,~~~~
 \log (q/\sigma) = { \beta \mu \over \tanh \tilde \gamma } 
 \eea

 It should be noted that Eq. (\ref{BCTh}) determine the inverse temperature $\beta$ and the parameters $\alpha$, and $\gamma$ in the solution as the functions of the parameter $\sigma$. However, it turns out that the system temperature $T_s(\sigma)=\beta^{-1}(\sigma)$ is not a monotonic function of $\sigma$. Due to the non-monotonicity of $T_s$, there is a temperature window $T_1<T_s<T_2$ in which there are three solutions with different energy $E$ for a given temperature. They correspond to three saddle points of the free energy, including two minima and one saddle point. This point will be explicitly demonstrated on the free energy landscape in the next section. 

 From Eq.(\ref{partfunc}), one can compute the free energy, the energy and also the entropy of the system, which are given by \cite{Maldacena:2018lmt}
 \bea
 {\tilde F_{on-shell}} &\equiv& { F \over N} = - {  \hat \mu \over q^2 } \left[  { q \over 2 } - 1 + { 1 \over \tanh \gamma \tanh \tilde \gamma } + \log { \sinh \gamma \over \cosh \tilde \gamma } + { \sigma \over \tanh \tilde \gamma } \right] + { \sigma \over q }
 \cr
 {\tilde E} &\equiv& { E \over N } = { \hat \mu \over q^2 } \left[ - { q \over 2 } + 1 - { 1 \over \tanh \gamma \tanh \tilde \gamma } - \log { \sinh \gamma \over \cosh \tilde \gamma }
 \right]
 \cr 
 {\tilde S} &\equiv& { S \over N } = { \sigma \over q} \left[ 1 + \log { q\over \sigma } \right]\,
 \eea
 Where we have denote the free energy calculated from the partition function as the on-shell free energy ${\tilde F_{on-shell}}$.

\begin{figure}
  \centering
  \includegraphics[width=7cm]{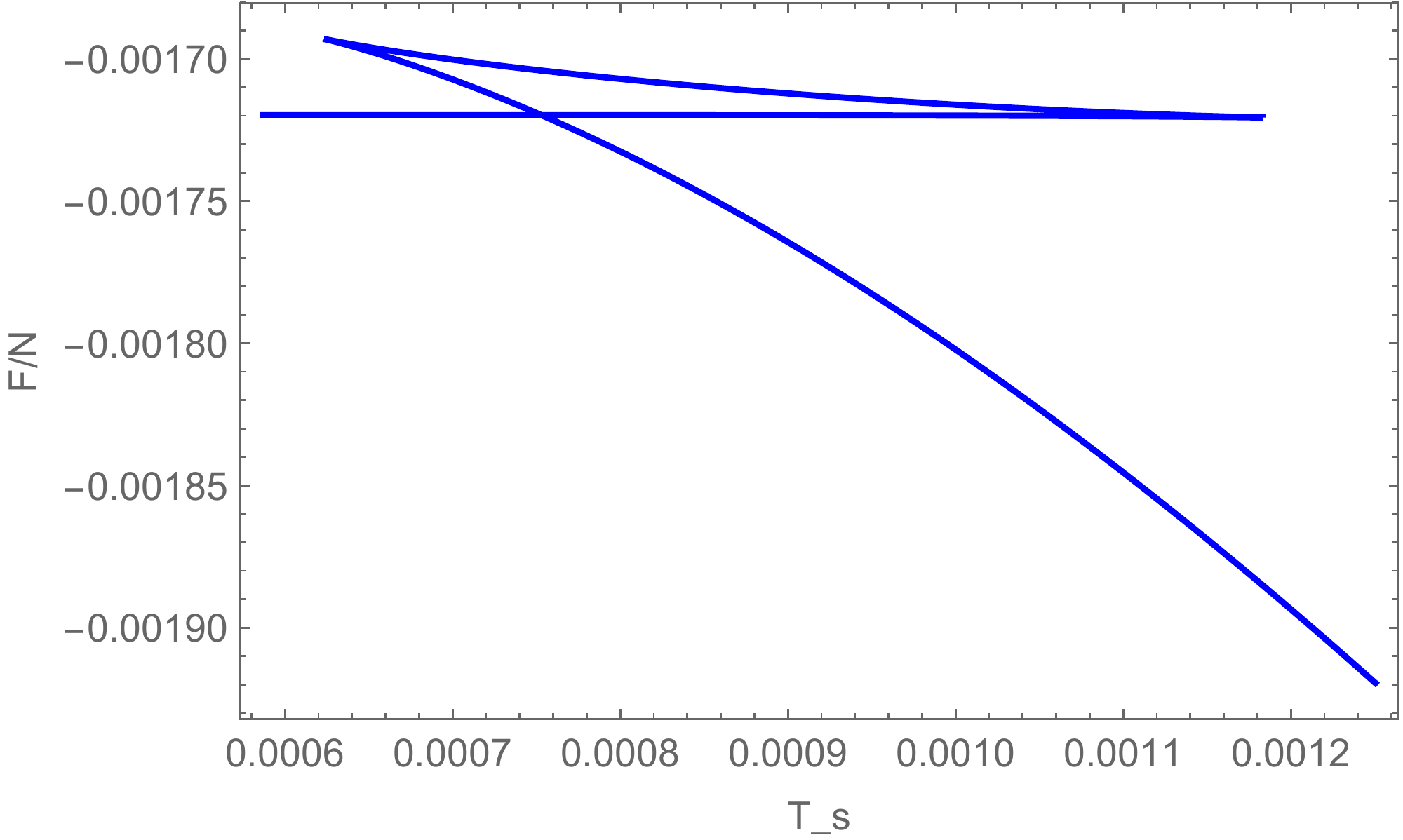}
  \caption{On-shell free energy ${\tilde F_{on-shell}}={ F \over N}$ as a function of the temperature $T_s$ of the system. In this plot, $q = 50$, ${\cal J} = 1$, and $\hat \mu = 0.1$. There is a first order phase transition at the temperature $T_s=0.000753$. }
  \label{onshell_Free_energy}
\end{figure}

\begin{figure}
  \centering
  \includegraphics[width=5cm]{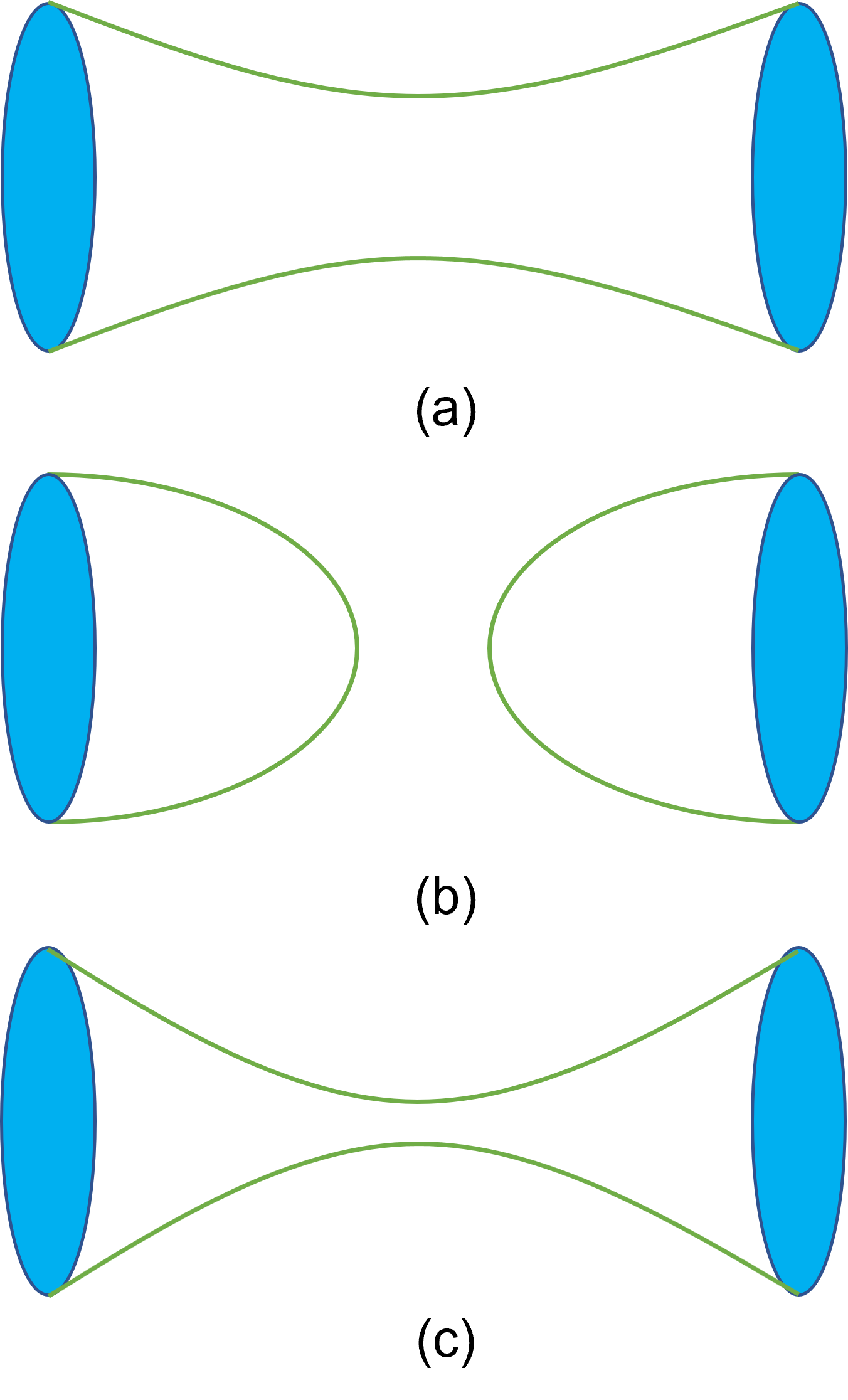}
  \caption{(a) stable wormhole phase in the low temperature; (b) two black hole phase in the high temperature; (c) unstable wormhole phase in the low temperature. Note that these pictorial representations are in the euclidean signature. }
  \label{wormhole_black_hole_pic}
\end{figure}

The on-shell free energy of the two coupled SYK model as the function of the system temperature $T_s$ is plotted in Figure \ref{onshell_Free_energy}. When $0.000623<T_s<0.00118$, there are three phases \cite{Maldacena:2018lmt}, which are the stable wormhole phase, the unstable wormhole phase and the two black hole phase as depicted in Figure \ref{wormhole_black_hole_pic}. The stable phase is the one that has the lowest on-shell free energy.

\section{Free energy landscape}\label{landscape}

We are interested in the emergence of wormhole and two black hole phases and the corresponding first order phase transition on the free energy landscape. In the formalism of the free energy landscape \cite{FSW,FW,NG,JW}, the Gibbs free energy is defined as the continuous function of the order parameter or reaction coordinate of the system. The order parameter or reaction coordinate may be considered as the coarse-grained description which captures the essential characteristics and the microscopic degree of freedom of the system \cite{Li:2020nsy,Li:2020khm,Li:2021vdp,Wei:2020rcd,Li:2020spm,Wei:2021bwy,Cai:2021sag,Lan:2021crt,Li:2021zep,Yang:2021nwd,Mo:2021jff,Kumara:2021hlt}.  

\begin{figure}
  \centering
  \includegraphics[width=7cm]{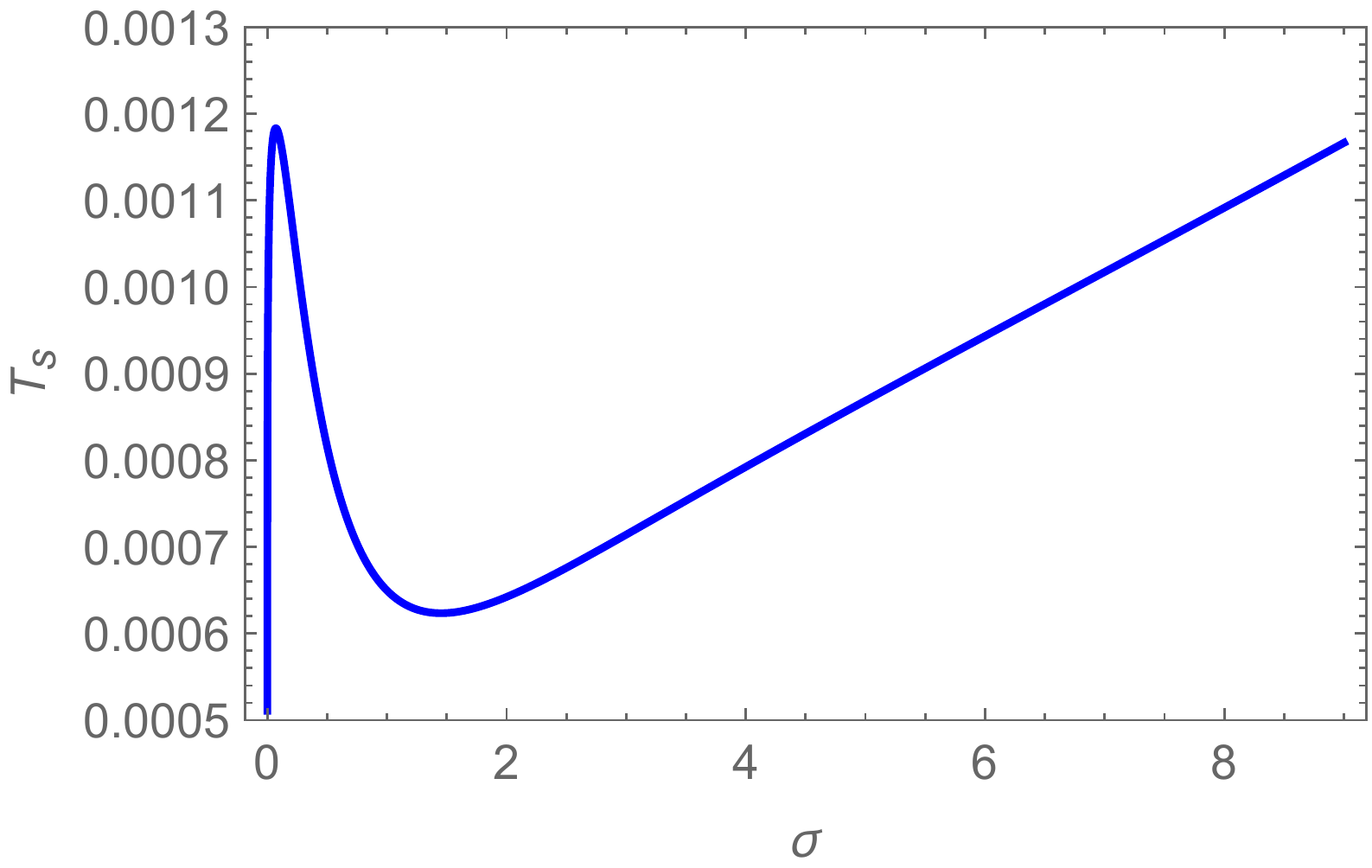}
  \caption{The system temperature $T_s$ as the function of the parameter $\sigma$. The non-monotonicity is shown explicitly. In this plot, $q = 50$, ${\cal J} = 1$, and $\hat \mu = 0.1$. }
  \label{T_vs_sigma}
\end{figure}

Let us firstly consider the system temperature $T_s=1/\beta$ as the function of the parameter $\sigma$. In Figure \ref{T_vs_sigma}, it is shown that the system temperature $T_s$ is not a monotonic function of $\sigma$. This is similar to the relationships between the temperature and black hole radius for the Hawking-Page transition \cite{Li:2020khm} or the small/big RNAdS black hole phase transition \cite{Li:2020nsy}, where the black hole radii are treated as the order parameters. Therefore, the non-monotonic relationship between the temperature and the parameter $\sigma$ prompts us the possibility of selecting $\sigma$ as the order parameter.

\begin{figure}
  \centering
  \includegraphics[width=7cm]{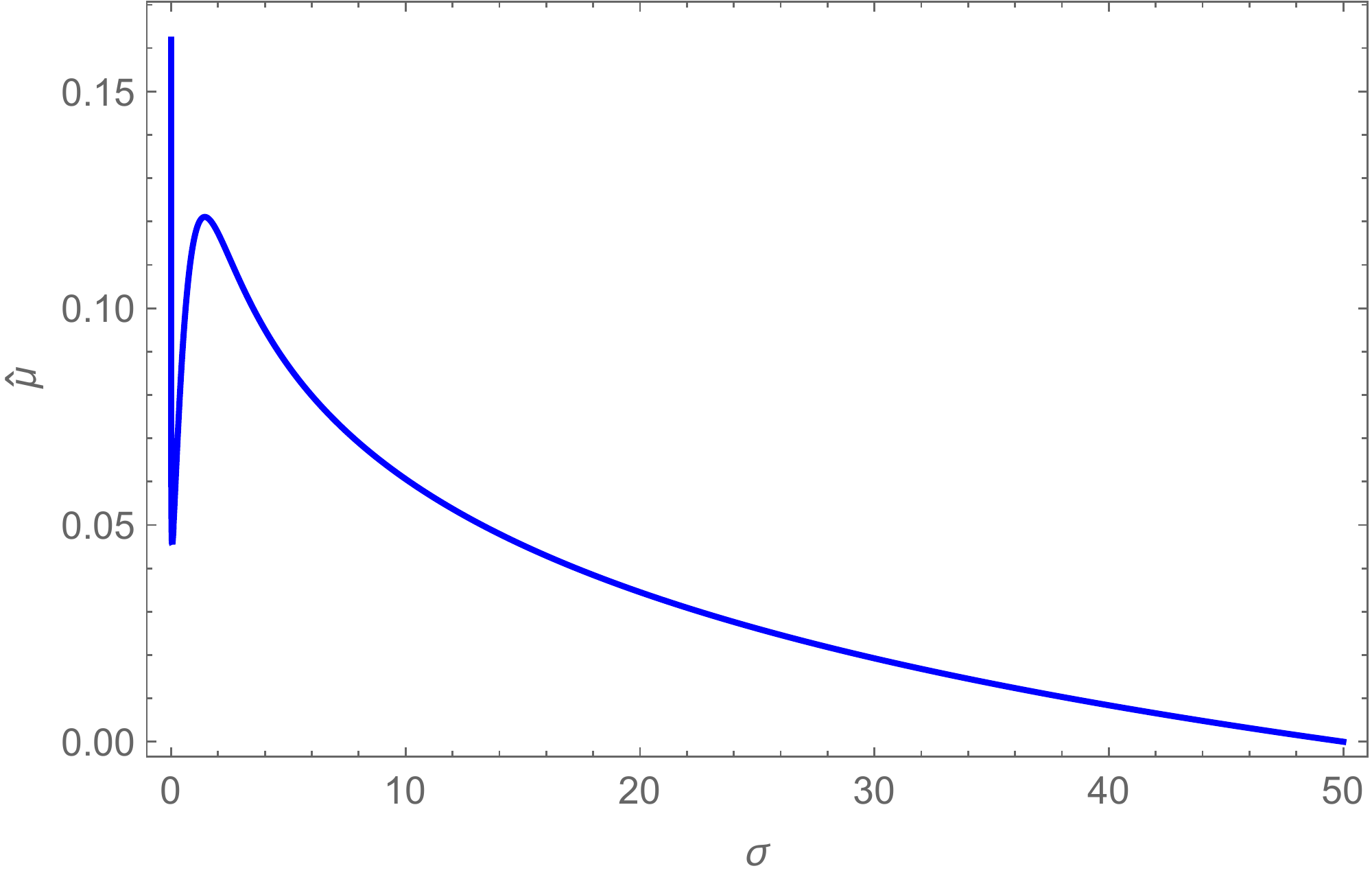}
  \caption{The coupling constant $\mu$ as the function of the parameter $\sigma$. The non-monotonicity is shown explicitly. In this plot, $q = 50$, ${\cal J} = 1$, and $T_s=0.000753$. }
  \label{mu_vs_sigma}
\end{figure}

Then we also consider the relationship between the coupling constant $\hat{\mu}$ and the parameter $\sigma$ in Figure \ref{mu_vs_sigma}. It is shown that their functional relationship is also non-monotonic. However, when $\sigma$ approaches zero, the coupling constant $\hat{\mu}$ becomes large, and when $\sigma$ becomes large, $\hat{\mu}$ approaches zero. Therefore, $\sigma$ represents the coupling strength of the two SYK models to a certain degree. In fact, the emergence of the three phases (the stable wormhole phase, the unstable wormhole phase and the two black hole phase in Figure \ref{wormhole_black_hole_pic} should reflect the coupling strength of the two coupled SYK models \cite{Maldacena:2018lmt}. This observation complemented with the non-monotonic relationship between $T_s$ and $\sigma$ indicates that the parameter $\sigma$ can be selected to be the order parameter.

We select $\sigma$ as the order parameter of the system. $\sigma$ should be considered as the independent variable. We consider the canonical ensemble at the specific temperature $T$. The temperature $T$ should be considered as the ensemble temperature, and is not determined by the order parameter $\sigma$. In order to construct the free energy landscape, we need to specify every state in the ensemble a Gibbs free energy. In other words, we should define the off-shell free energy as a function of the order parameter and the ensemble temperature. Recall that the on-shell free energy is given by the thermodynamic relationship $F=E-T_s S$, where $T_s$ is the temperature of the system, which is different from the ensemble temperature $T$. Now we generalized this definition to the off-shell free energy as 
\bea 
{\tilde F} &=& {\tilde E} - T {\tilde S}\nonumber\\
&=&
{ \hat \mu \over q^2 } \left[ - { q \over 2 } + 1 - { 1 \over \tanh \gamma \tanh \tilde \gamma } - \log { \sinh \gamma \over \cosh \tilde \gamma }
 \right] -{ \sigma \over q} \left[ 1 + \log { q\over \sigma } \right] T
\eea
where $T$ denotes the ensemble temperature. Without special remark, $T$ denotes the ensemble temperature but not the temperature of the system in the following. In the above expression, $\tilde \gamma$ and $\gamma$ are considered as the functions of $\sigma$ by the following relations 
\bea
\tilde \gamma = \gamma + \sigma ~,~~~~ \hat \mu = 2 {\cal J} \sinh \gamma \tanh \tilde \gamma \;.
\eea

\begin{figure}
  \centering
  \includegraphics[width=7cm]{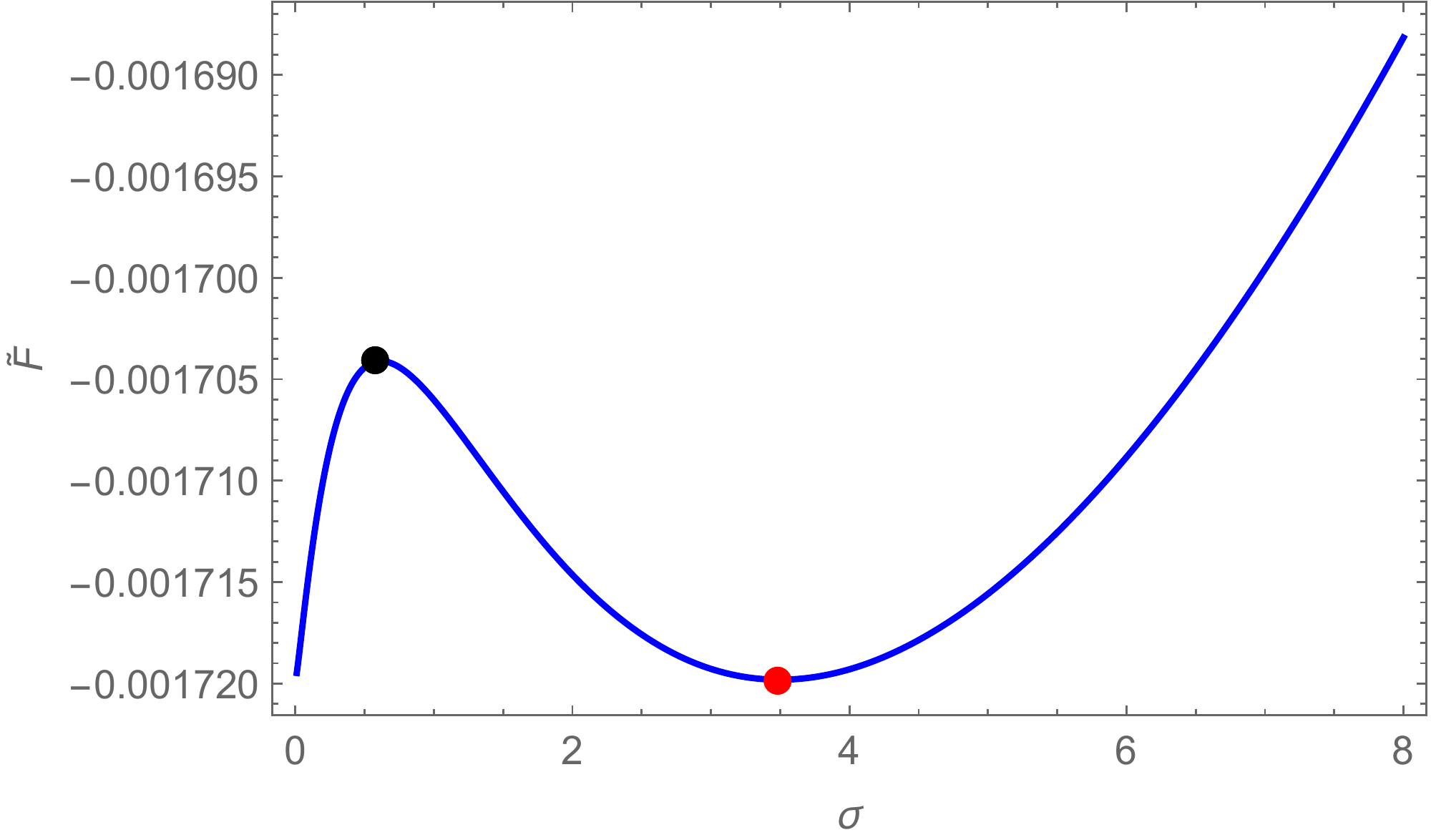}
  \includegraphics[width=7cm]{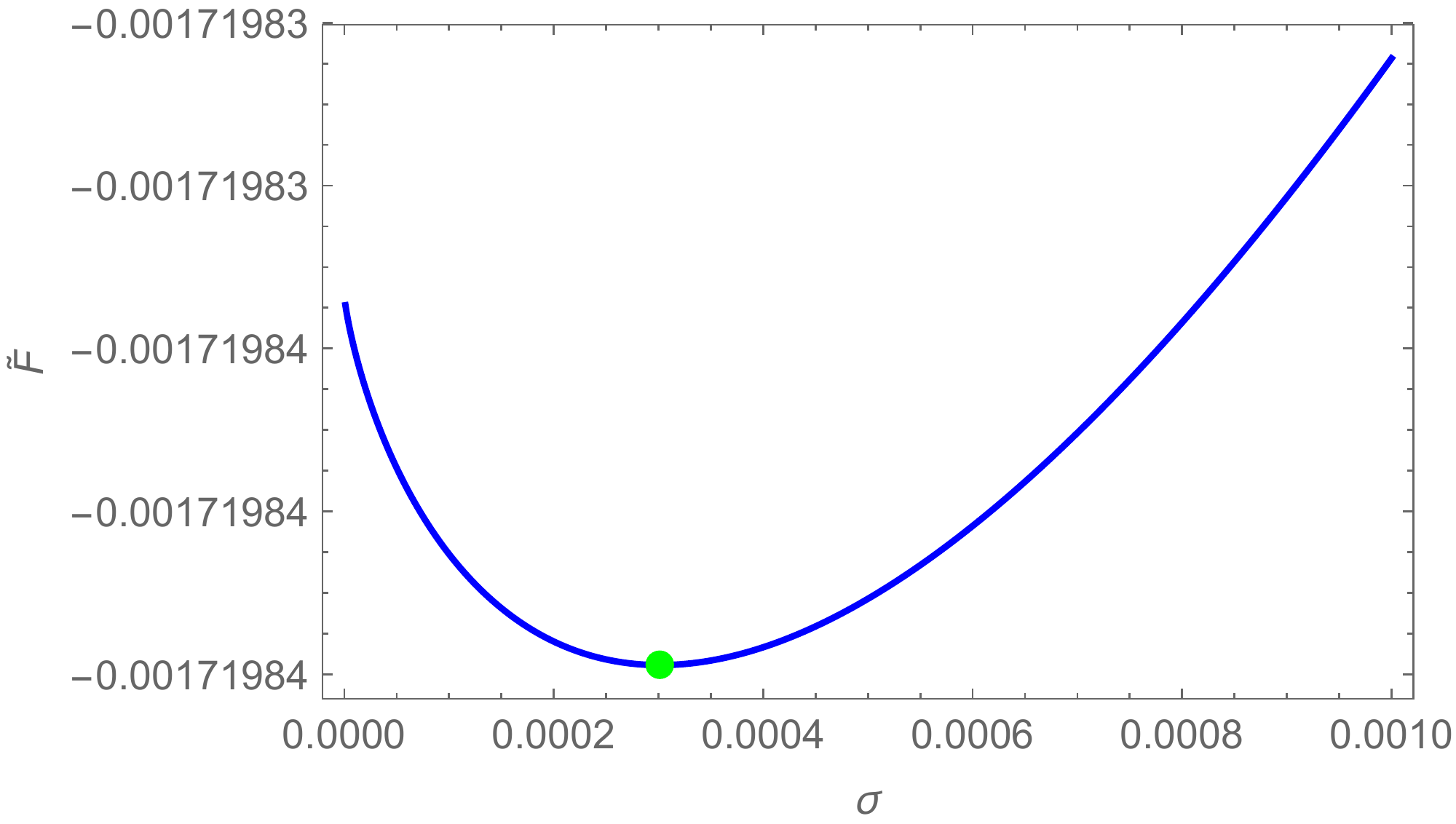}
  \caption{Free energy landscape, i.e. the off-shell free energy as the function of the order parameter $\sigma$. The right panel is the zoom in picture of the left panel. It is shown that the landscape has the shape of the double well and three saddle points representing the wormhole phases and two black hole phase. In this plot, $q = 50$, ${\cal J} = 1$, $\hat \mu = 0.1$, and $T=0.000753$. }
  \label{Free_energy_landscape}
\end{figure}

The free energy landscape, which was a visualized image of the functional relation between the off-shell free energy and the order parameter, is plotted in Figure \ref{Free_energy_landscape}. The right panel is the zooming in on the picture of the left panel. It is shown that the free energy landscape has the shape of the double well. This is different from the single well of Hawking-Page transition but similar to the double well shape of the free energy landscape of the small/large RNAdS black hole phase transition. Therefore, the first order phase transition should be van der Waals type. The three saddle points are displayed by the red, the green and the black dots correspondingly. According to the discussion above, the red, the green and the black dots in the free energy landscape corresponds to two black hole phase, stable wormhole phase and unstable wormhole phase respectively. For latter convenience, we denote the order parameters of the three phases as $\sigma_s$ for the stable wormhole phase, $\sigma_m$ for the unstable wormhole phase, and $\sigma_l$ for the two black hole phase. The ensemble temperature for the plots in Figure \ref{Free_energy_landscape} is taken as the phase transition temperature $T_{ph}=0.000753$, in which case the left and right wells representing the stable wormhole phase and the two black hole phase have the same depths.

\begin{figure}
  \centering
  \includegraphics[width=7cm]{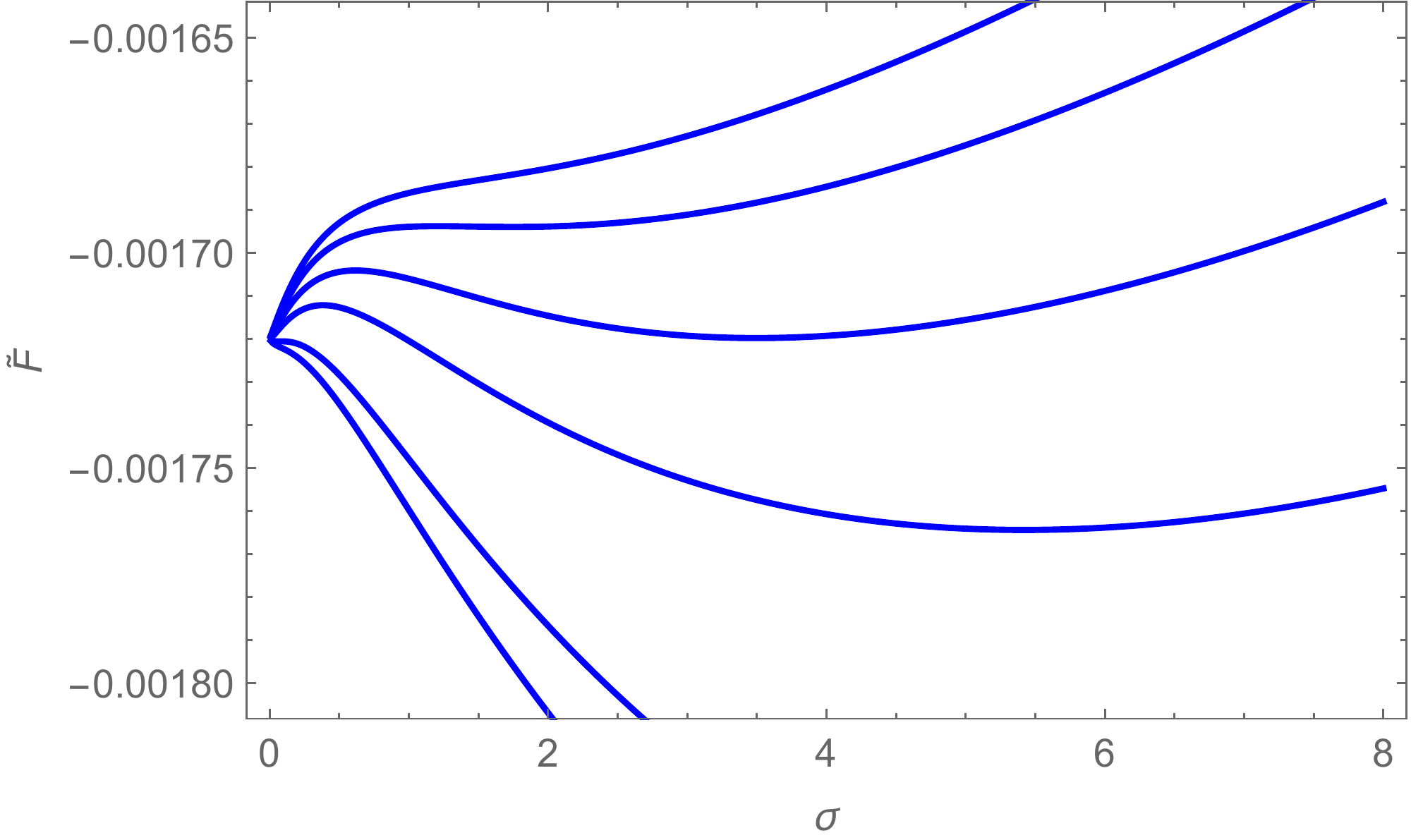}
  \includegraphics[width=7cm]{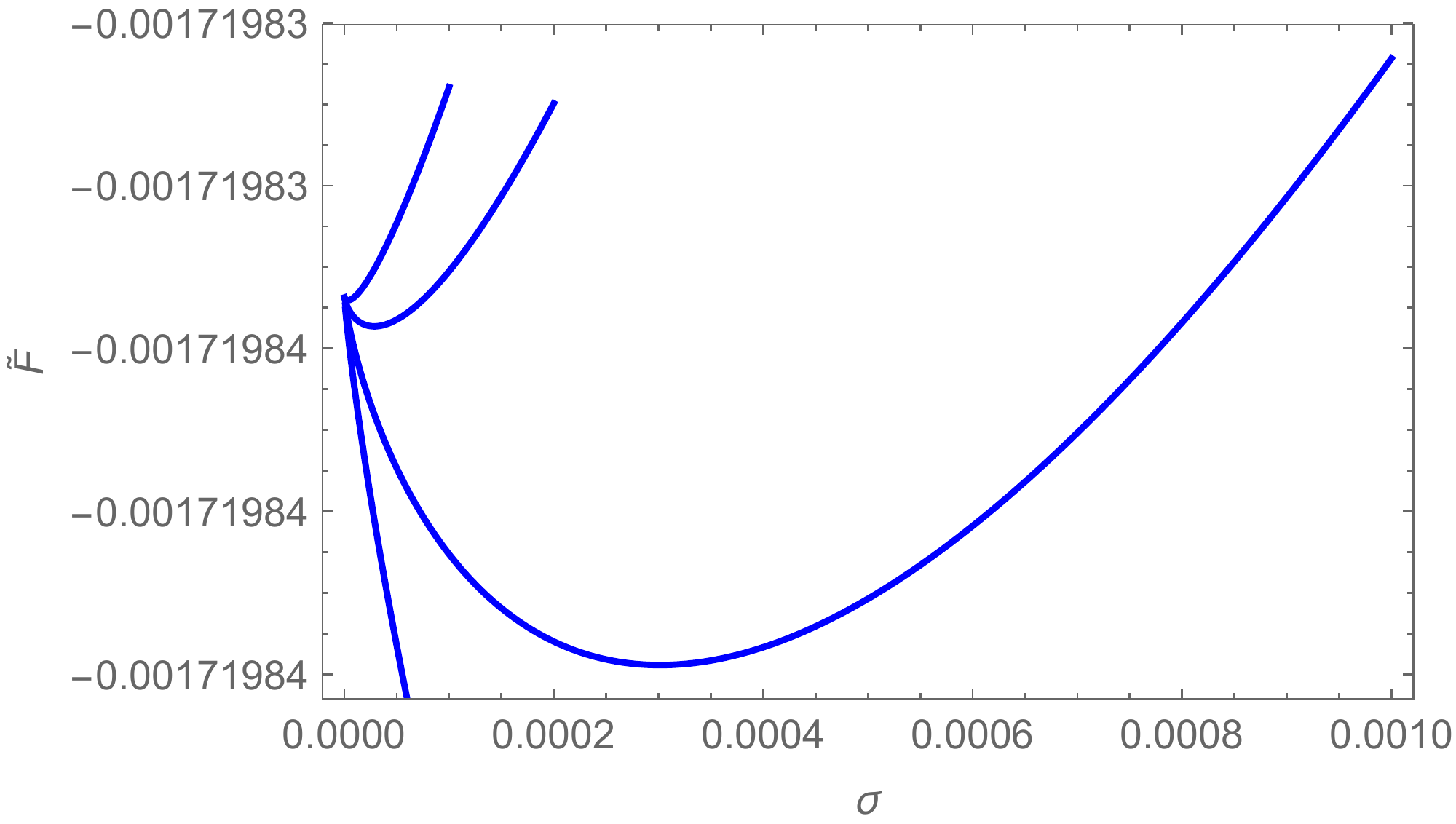}
  \caption{Free energy landscape at different ensemble temperatures. The right panel is the zoom in picture of the left panel. In the left panel, the ensemble temperatures are $0.00055, 0.00063, 0.000753, 0.0009, 0.00118$, and  $0.0013$ from the top to bottom. In the right panel, the ensemble temperatures are $0.00055, 0.00063, 0.000753$, and $0.0009$ from the top to bottom. In this plot, $q = 50$, ${\cal J} = 1$, and $\hat \mu = 0.1$.}
  \label{Free_energy_landscape_diffT}
\end{figure}

In Figure \ref{Free_energy_landscape_diffT}, we show the free energy landscapes at different ensemble temperatures. For $\hat\mu=0.1$, there are two temperatures $T_1=0.00063$ and $T_2=0.00118$. We find that when $T_1<T<T_2$, three phases (two stable and one unstable phases) coexist and the free energy landscape has the shape of double well. In particular, when $T_1<T<T_{ph}$, the depth of the left well representing the stable wormhole phase is lower than the depth of the right well representing the two black hole phase, which implies that the wormhole phase is thermodynamically favored. When $T_{ph}<T<T_{2}$, the conclusion is exchanged and the two black hole phase is favored. When $T<T_1$ or $T>T_2$, the free energy landscape has the shape of the single well. From Figure \ref{Free_energy_landscape_diffT}, we also find that the depth of the right well changes significantly when varying the temperature, while the depth of the left well is insensitive.

\begin{figure}
  \centering
  \includegraphics[width=7cm]{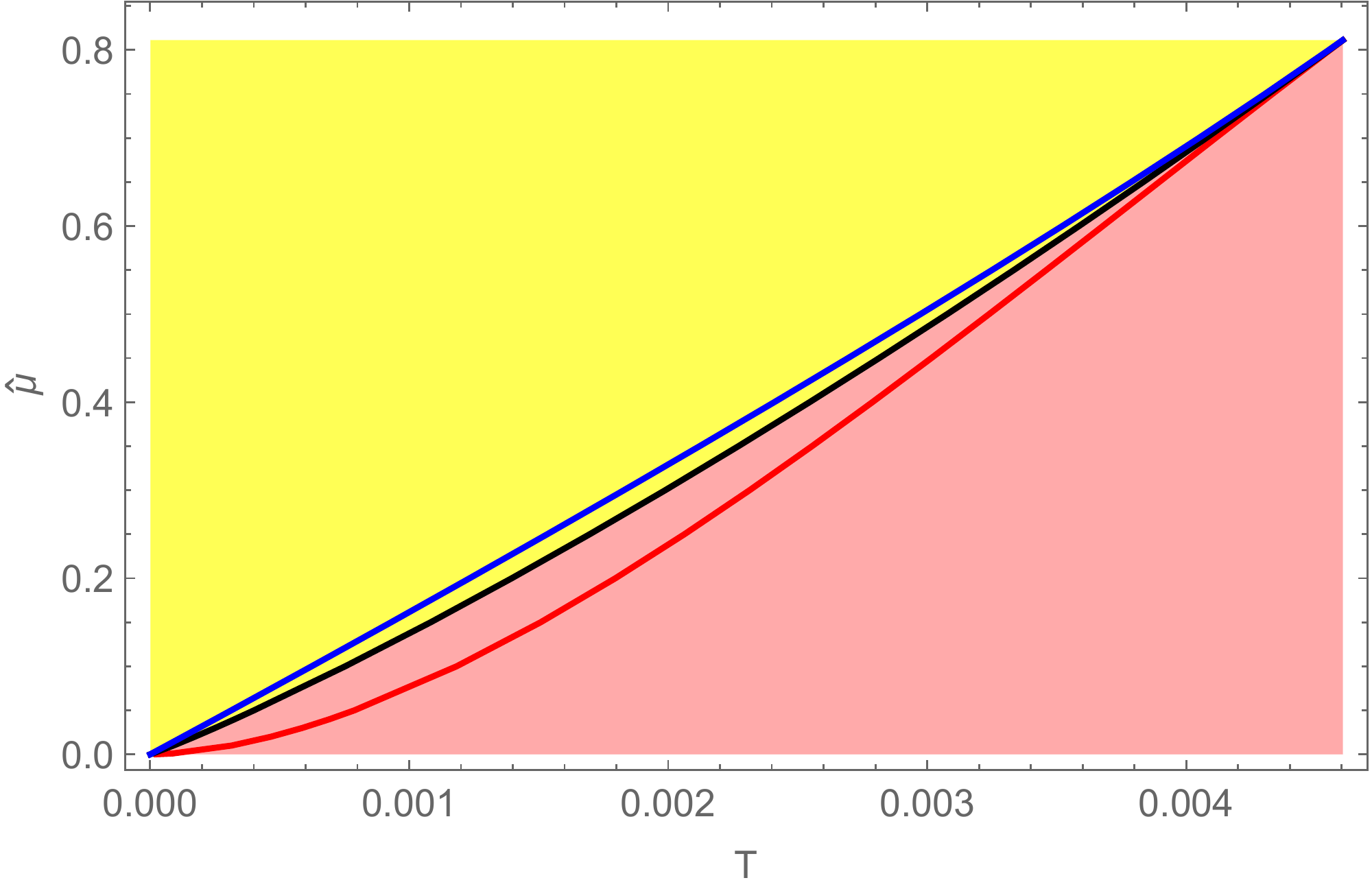}
  \caption{Phase diagram in $\hat\mu-T$ plane. In this plot, $q = 50$ and ${\cal J} = 1$.}
  \label{Phase_diagram}
\end{figure}

In order to confirm the argument of the phase transition type, we further calculate the phase diagram of the system, which is presented in Figure \ref{Phase_diagram}. 
The stable wormhole phase is thermodynamically favored in the yellow region, while the two black hole phase is thermodynamically favored in the pink region. For specific coupling constant $\hat\mu$, there is a temperature window $T_1<T<T_2$ in which three phases coexist and the free energy landscape has the shape of double well. Varying the parameter $\hat\mu$, we can get $T_1$ and $T_2$ as the function of $\hat\mu$. These two curves $T_1(\hat\mu)$ and $T_2(\hat\mu)$ are plotted in blue and red in Figure \ref{Phase_diagram}. There is also a phase transition temperature, which is plotted in black. The phase transition temperature is determined by equal free energies of the two basins. The black line represents the coexistence curve that the stable wormhole phase and the two black hole phase coexist in equal probability. When $\hat\mu\rightarrow 0$ or $\hat\mu$ increases, the temperature window closes. It is shown there is a critical point as in the gas-liquid phase transition of van der Waals fluid. At the critical point, $\hat\mu=0.811$, and $T=0.0046$. This behavior is analogous to the van der Waal type phase transition in thermodynamics.   

\section{Kinetics of Phase Transition on the free energy landscape}\label{kinetics}

\subsection{Fokker-Planck equation}

We discuss the effective theory of the stochastic dynamics of the phase transition. Note that the generation of the fluctuating black hole is completely stochastic and the dynamical process is assumed to have the coarse-grained description by using the order parameter. In this section, we will study the kinetics of the phase transition based on the free energy landscape. As shown in the last section, we know that the free energy landscape as a function of the order parameter $\sigma$ exhibits double basin shape when the temperature lies in the region $[T_1, T_2]$.

We consider the case where are a large number of states in a thermodynamic ensemble in which one or a group of them can represent the wormholes or the two black hole phases during the state transition process. The probability distribution of the states (on-shell solutions as well as the off-shell solutions on the free energy landscape) are denoted by $\rho(\sigma)$. The stochastic kinetics of states under the thermal fluctuation can be described by the probabilistic Fokker-Planck equation, which on the free energy landscape is explicitly given by \cite{Li:2020nsy,Li:2020khm,Li:2021vdp,Wei:2020rcd,Li:2020spm,Wei:2021bwy,Cai:2021sag,Lan:2021crt,Li:2021zep,Yang:2021nwd,Mo:2021jff,Kumara:2021hlt}
\begin{eqnarray}\label{FPeq}
\frac{\partial \rho(\sigma,t)}{\partial t}=D \frac{\partial}{\partial \sigma}\left\{
e^{-\tilde \beta \tilde F(\sigma)}\frac{\partial}{\partial \sigma}\left[e^{\tilde \beta \tilde F(\sigma)}\rho(\sigma,t)\right]
\right\}\;.
\end{eqnarray}
In the above equation, the diffusion coefficient $D$ is given by $D=kT/\zeta$ with $k$ being the Boltzman constant and $\zeta$ being dissipation coefficient. Without the loss of generality, we will take $k=\zeta=1$ in the following. Note that $\tilde\beta=1/k T$ is the inverse ensemble temperature. Note that $\tilde F(\sigma)$ is the off-shell free energy as a function of the order parameter $\sigma$ modulated by the ensemble temperature $T$.

\subsection{Analytical results of the kinetic time and its fluctuation}

We now study the kinetics through the mean first passage time for the state switching process.
Our purpose is to find out the time that it takes for a state starting from one local(global) stable phase and ending to another global(local) stable phase. Problems of this sort can be solved by the first passage time. In general, first passage time is defined as the time required for a state from the local(global) stable phase (either stable wormhole or two black hole described by the local(global) minimum of Gibbs free energy) to reach the transition state (an unstable intermediate wormhole phase determined by the maximum of Gibbs free energy or barrier in the present case) for the first time. Since we are considering a stochastic process under thermal fluctuation, the first passage time is a random variable. Thus, we are interested in the mean first passage time particularly. The mean first passage time quantifies an average timescale for a stochastic event of switching to first occur.

According to the discussion in \cite{Li:2020khm}, 
one can obtain the analytical integration expression for the mean first passage time from the stable wormhole phase to the unstable wormhole phase as
\begin{eqnarray}\label{MFPTexp1}
\langle t \rangle&=&\frac{1}{D}\int_{\sigma_s}^{\sigma_m}d\sigma \int_{0}^{\sigma}d\sigma'  e^{\tilde \beta  \left(\tilde F(\sigma)-\tilde F(\sigma')\right)}\;.
\end{eqnarray}
Note that $\sigma_s$ and $\sigma_m$ are the order parameters for the stable wormhole phase and the unstable wormhole phase. In deriving this expression, we have imposed the reflecting boundary at $\sigma=0$.

The analytical integral expressions for the mean first passage time from the two black hole phase to the unstable wormhole phase can also be derived similarly, which is given as follows
\begin{eqnarray}\label{MFPTexp2}
\langle t \rangle&=&\frac{1}{D}\int_{\sigma_m}^{\sigma_l}d\sigma \int_{\sigma}^{+\infty}d\sigma'  e^{\tilde \beta  \left(\tilde F(\sigma)-\tilde F(\sigma')\right)}\;.
\end{eqnarray}
Note that $\sigma_l$ denotes the order parameter of the two black hole phase. With these expressions, we can compute mean first passage time via numerical integration directly without concern on the time distribution.

We can also derive the following analytical integration expression of $\langle t^2 \rangle$ for the state transition process from the stable wormhole phase to the unstable wormhole phase that can be used to compute the kinetic fluctuations of the first passage time 
\begin{eqnarray}\label{MFPT2exp1}
\langle t^2 \rangle&=&\frac{2}{D^2}\int_{\sigma_s}^{\sigma_m}d\sigma \int_{0}^{\sigma}d\sigma'
\int_{\sigma'}^{\sigma_m}dr'' \int_{0}^{\sigma''}d\sigma'''
e^{\tilde \beta  \left(\tilde F(\sigma)-\tilde F(\sigma')+\tilde F(\sigma'')-\tilde F(\sigma''')\right)}\;,
\end{eqnarray}
and the similar analytical integral expressions of $\langle t^2 \rangle$ for the state transition process from the two black hole phase to the unstable wormhole phase 
\begin{eqnarray}\label{MFPT2exp2}
\langle t^2 \rangle&=&\frac{2}{D^2}\int_{\sigma_m}^{\sigma_l}d\sigma \int_{\sigma}^{+\infty}d\sigma'
\int_{\sigma_m}^{\sigma'}d\sigma'' \int_{\sigma''}^{+\infty}d\sigma'''
e^{\tilde \beta  \left(\tilde F(\sigma)-\tilde F(\sigma')+\tilde F(\sigma'')-\tilde F(\sigma''')\right)}\;.
\end{eqnarray}

\subsection{Numerical results of the kinetic time and its fluctuation}

Firstly, we consider the kinetics and its fluctuation of the transition process from the stable wormhole phase (the green dot on the free energy landscape in Figure \ref{Free_energy_landscape}) to the unstable wormhole phase (black dot on the free energy landscape). We have assumed that this switching process dominates the phase transition process from the stable wormhole phase to the two black hole phase.

In general, the kinetics of the transition process can be described by two quantities, i.e. the mean first passage time and its fluctuation. Here, we mainly consider the relative fluctuation, which is defined as $\left(\langle t^2 \rangle-\langle t \rangle^2\right)/\langle t^2 \rangle$. According to the analytical integral expressions for $\langle t \rangle$ and $\langle t^2 \rangle$, we can obtain the numerical relationship between the mean first passage time and the ensemble temperature as well as the dependence relation between the relative fluctuation and the ensemble temperature.

\begin{figure}
  \centering
  \includegraphics[width=7cm]{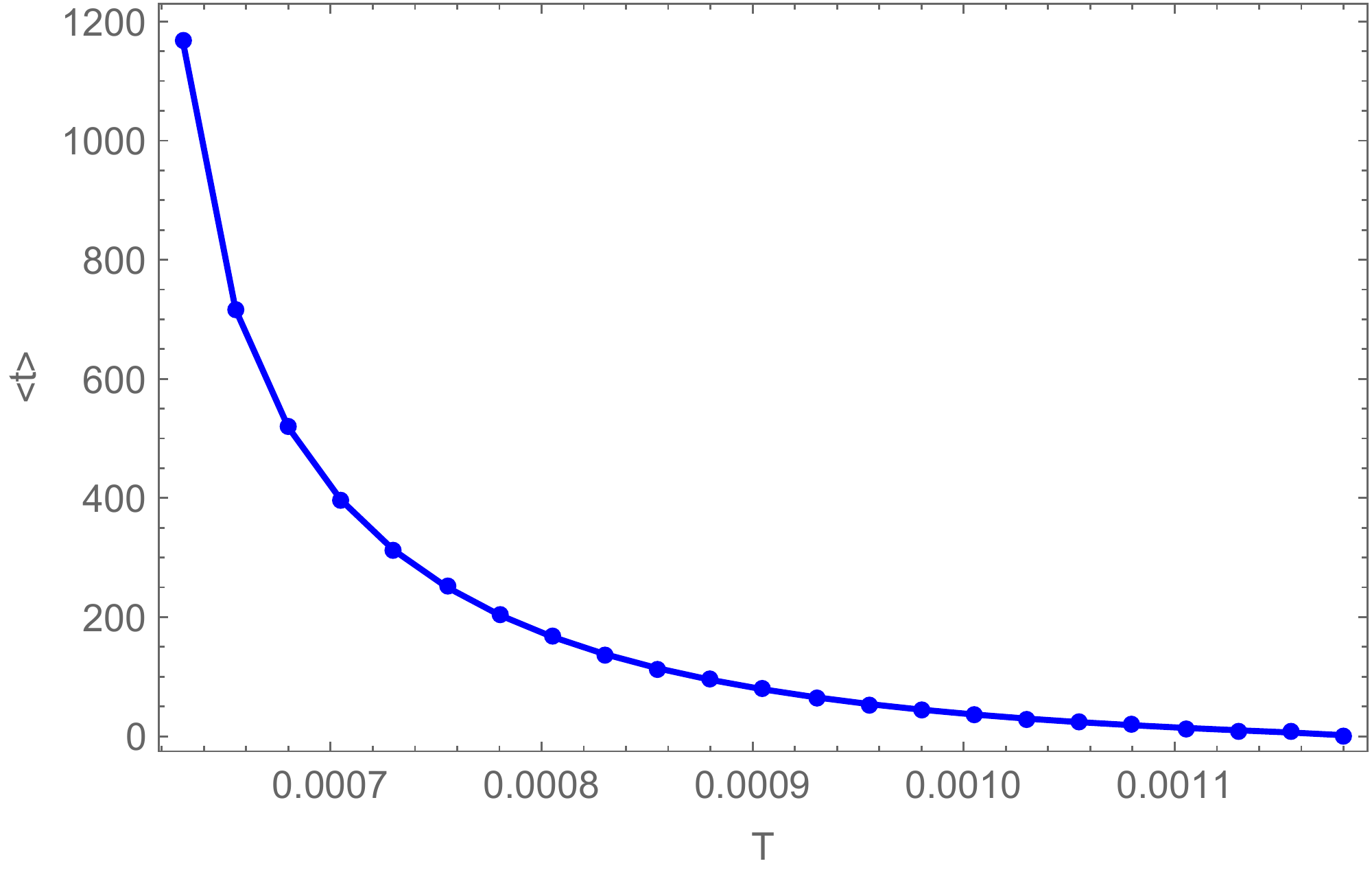}
  \includegraphics[width=7cm]{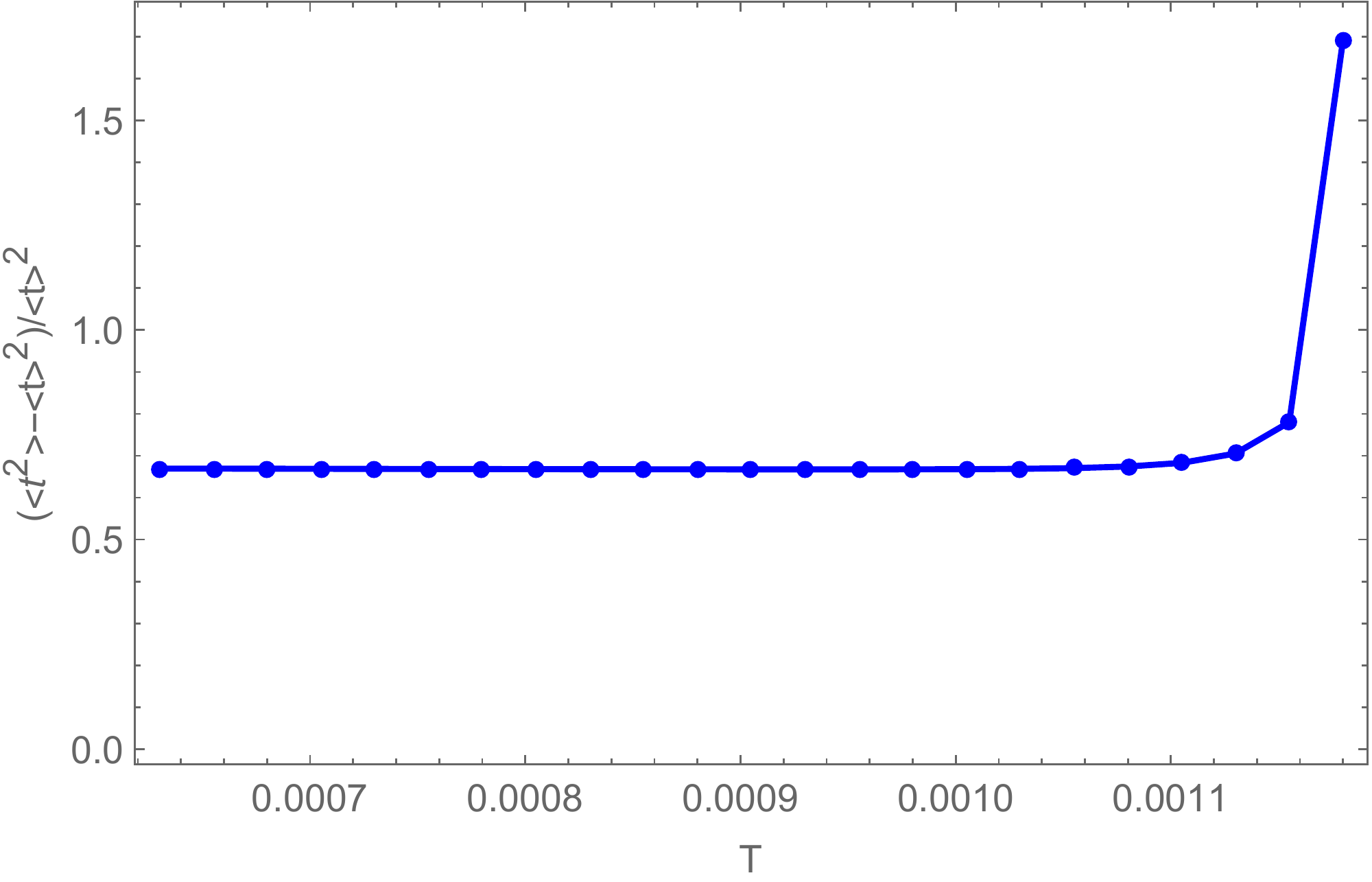}
  \caption{Mean first passage time (left panel) and the relative fluctuation (right panel) of the transition process from the stable wormhole phase to the unstable wormhole phase. In this plot, $q = 50$, ${\cal J} = 1$, and $\hat \mu = 0.1$. }
  \label{MFPT_s_to_m}
\end{figure}

\begin{figure}
  \centering
  \includegraphics[width=7cm]{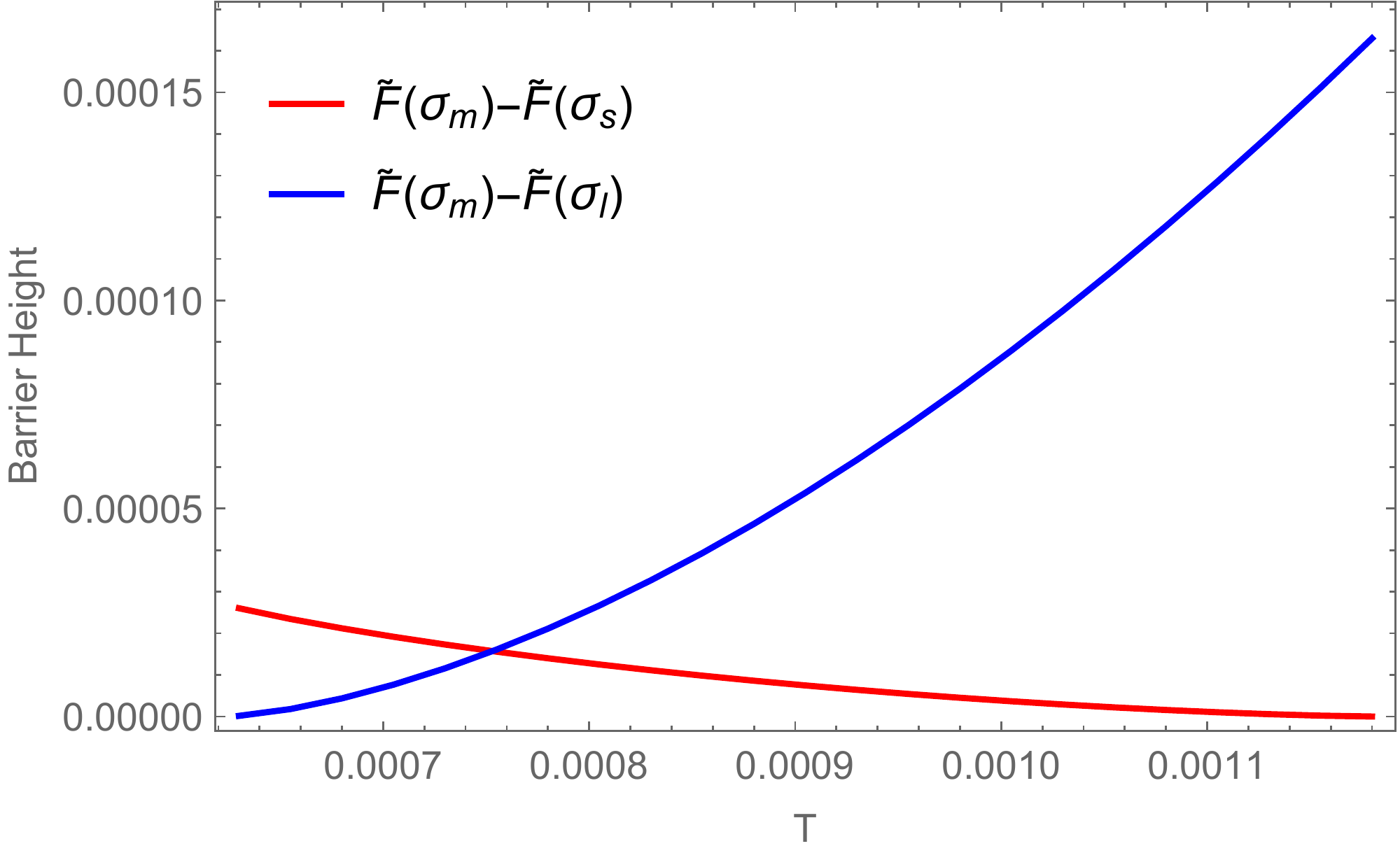}
  \caption{Barrier height (red line) between the stable wormhole phase (the green dot on the free energy landscape in Figure \ref{Free_energy_landscape}) and the unstable wormhole phase (black dot on the free energy landscape) and barrier height (blue line) between the two black hole phase (the red dot on the free energy landscape in Figure \ref{Free_energy_landscape}) and the unstable wormhole phase (black dot on the free energy landscape).}
  \label{barrier_height}
\end{figure}

\begin{figure}
  \centering
  \includegraphics[width=7cm]{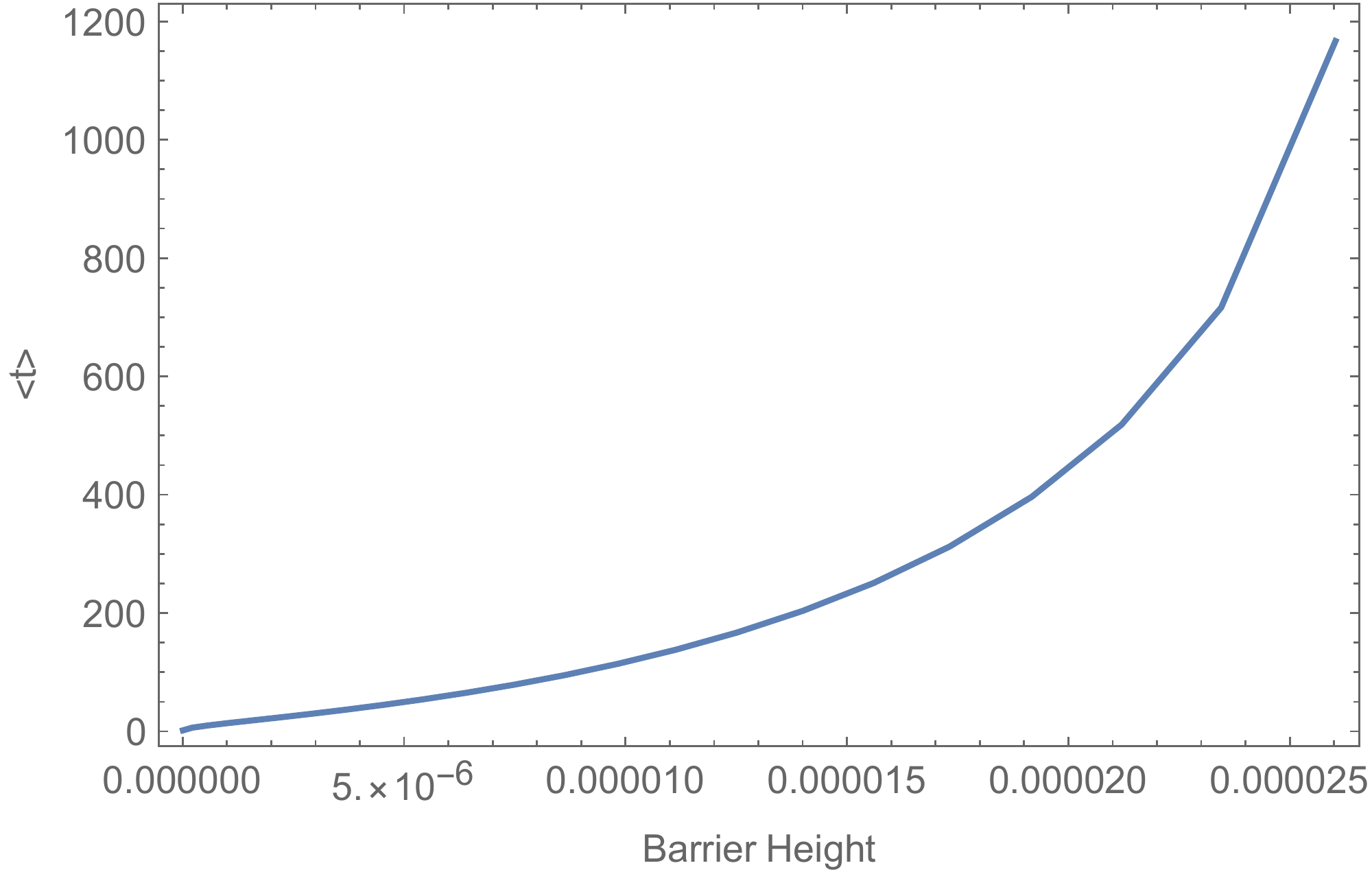}
  \caption{The correlation between the mean kinetic time $\langle t\rangle$ and the barrier height for the phase transition process from the stable wormhole phase to the unstable wormhole phase.}
  \label{barrier_height_vs_MFPT_stm}
\end{figure}

The numerical results of the mean first passage time $\langle t \rangle$ and the relative fluctuation for the transition process from the stable wormhole phase to the unstable wormhole phase are presented in Figure \ref{MFPT_s_to_m}. We are mainly interested in their temperature dependence. The temperature region is taken to be $[T_1, T_2]$, in which the double well shape of the free energy landscape emerges. It is shown that the mean first passage time for the transition process from the stable wormhole phase to the unstable wormhole phase is a monotonic decreasing function of temperature. The main reason of this behavior is that the kinetics is determined by the barrier height between the two phases on the free energy landscape. When increasing the temperature, the barrier height on the free energy landscape between the stable wormhole phase and the unstable wormhole phase becomes smaller as shown in Figure \ref{barrier_height}. In order to make this argument more clear, we have illustrated the correlation between the barrier height and the mean kinetic time $\langle t\rangle$ in Figure \ref{barrier_height_vs_MFPT_stm}. It is shown that kinetic time is a monotonous function of the barrier height. This illustrates how the free energy landscape topography in terms of the barrier height determines the kinetics of the phase transition. The second reason of the behavior of the kinetic time is due to the temperature, since the switching, which is the diffusion process caused by the thermal fluctuation, becomes more and more effective at higher temperatures.

The numerical results of the relative fluctuation of the kinetic time shows that the relative fluctuation is small in most of the temperature range while it becomes big when temperature approaches $T_2$. It is obvious that the relative fluctuations attain the maximum at the temperature where the mean first time is at its minimum. We believe that the behavior of the relative fluctuations of the first passage time is the consequence of the two elements mentioned above. As the temperature increases, the free energy barrier height becomes smaller as shown in Figure \ref{barrier_height} while the temperature becomes higher. The higher thermal fluctuations relative to the barrier height indicate that the thermal fluctuations will have more significant impacts on the kinetics and associated fluctuations than the free energy barrier height at high temperatures. Therefore, the larger thermal fluctuations relative to the free energy barrier height lead to larger relative fluctuations in kinetics at high temperatures. At last, it can be concluded that the effect of the temperature dominates.

\begin{figure}
  \centering
  \includegraphics[width=7cm]{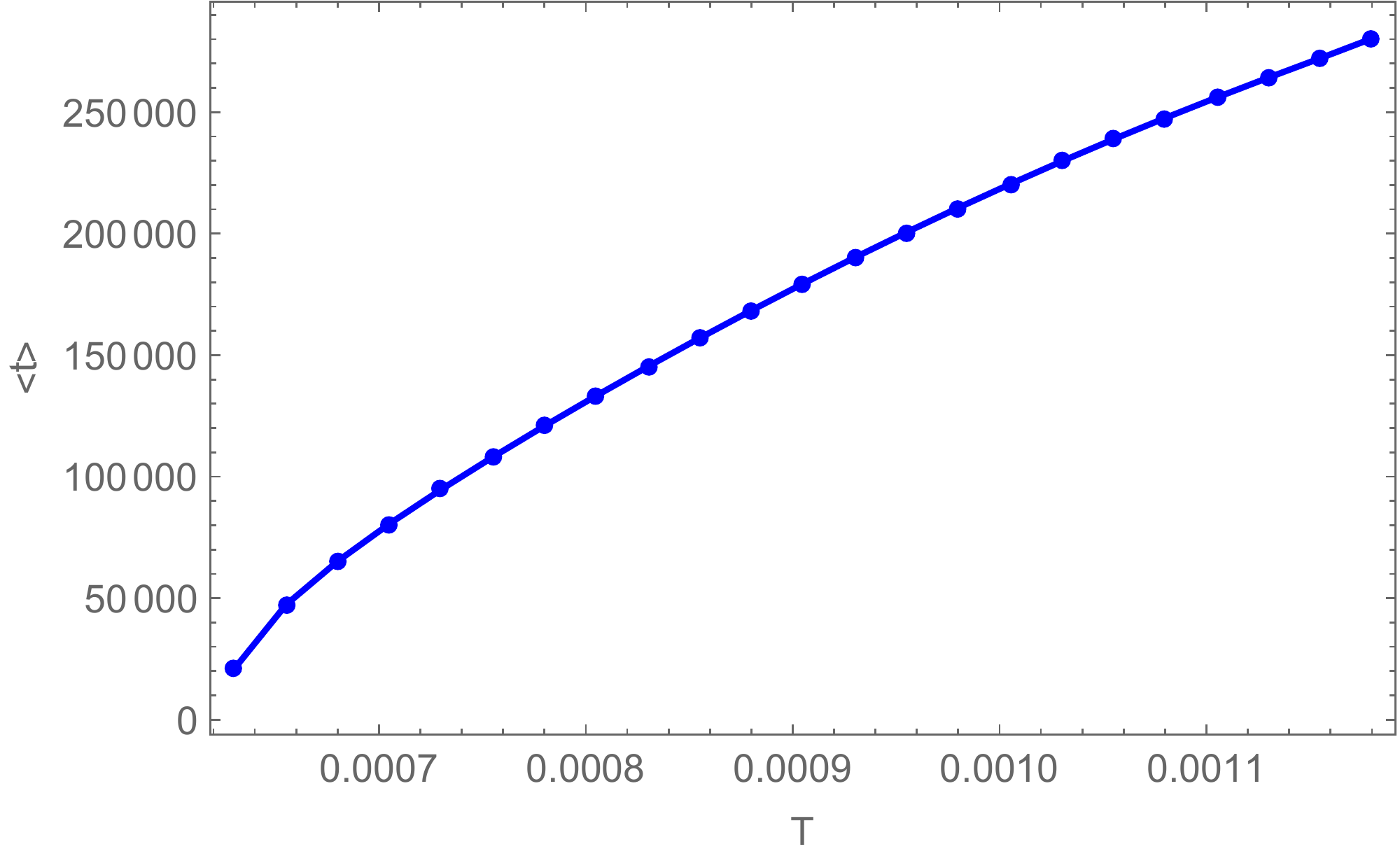}
  \includegraphics[width=7cm]{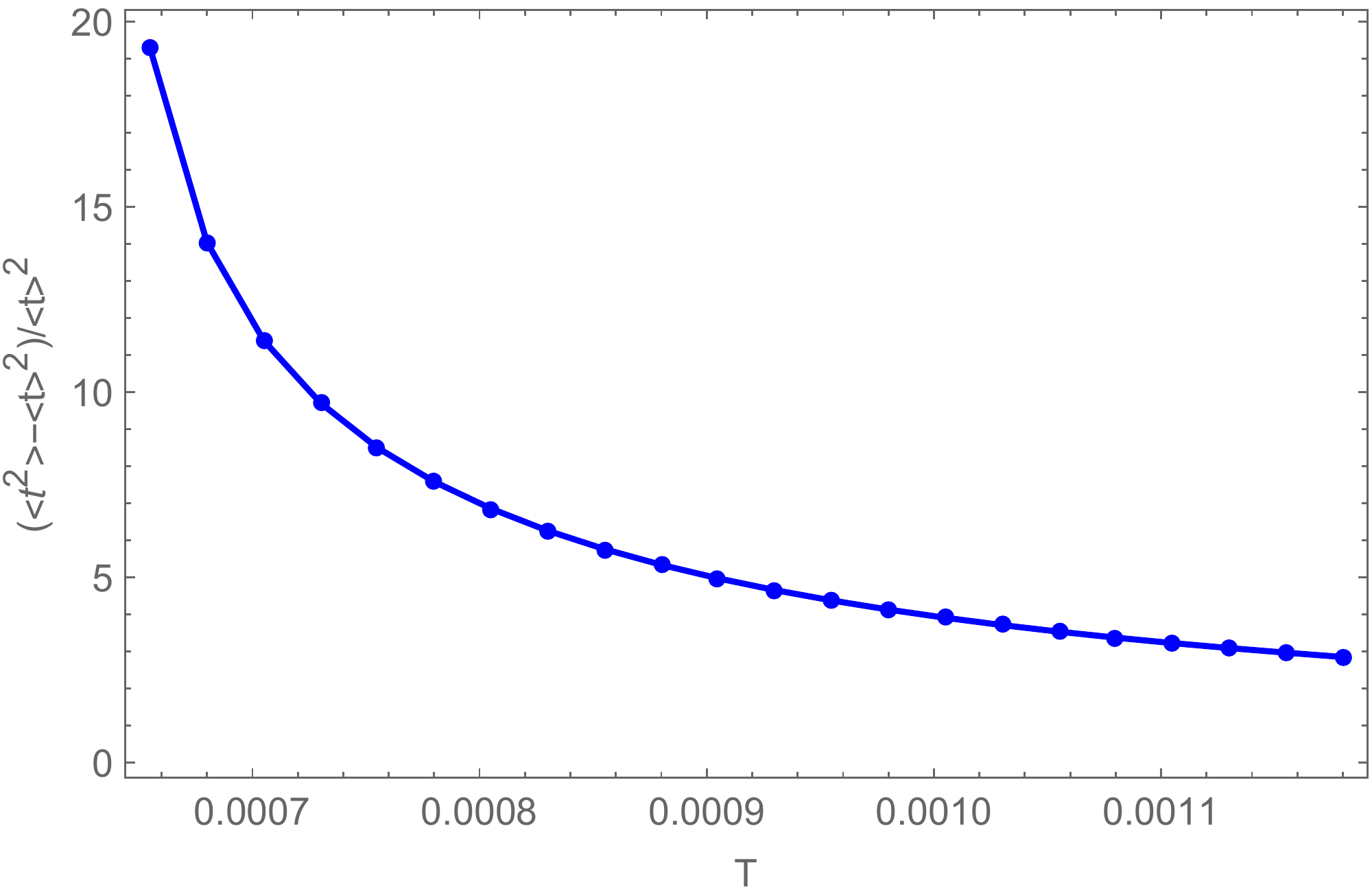}
  \caption{Mean first passage time (left panel) and the relative fluctuation (right panel) of the transition process from the two black hole phase to the unstable wormhole phase. In this plot, $q = 50$, ${\cal J} = 1$, and $\hat \mu = 0.1$. }
  \label{MFPT_l_to_m}
\end{figure}

\begin{figure}
  \centering
  \includegraphics[width=7cm]{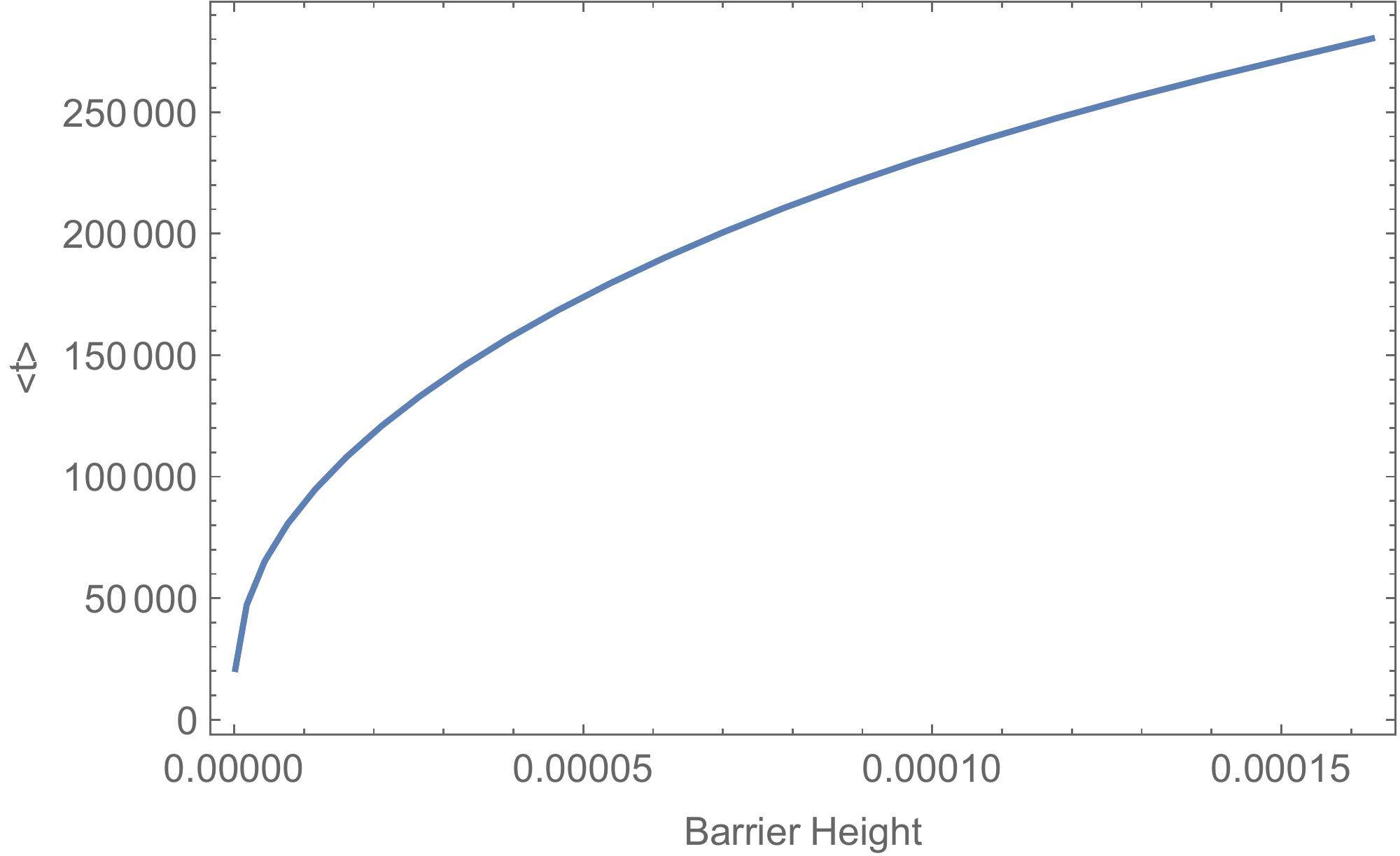}
  \caption{The correlation between the mean kinetic time $\langle t\rangle$ and the barrier height for the phase transition process from the two black hole phase to the unstable wormhole phase.}
  \label{barrier_height_vs_MFPT_ltm}
\end{figure}

Now, we consider the kinetics and its fluctuation of the transition process from the two black hole phase (the red dot on the free energy landscape in Figure \ref{Free_energy_landscape}) to the unstable wormhole phase (black dot on the free energy landscape). We also assume that this switching dominates the phase transition process from the two black hole phase to the stale wormhole phase. The numerical results of the mean first passage time $\langle t \rangle$ and the relative fluctuation for the transition process from the two black hole phase to the unstable wormhole phase are presented in Figure \ref{MFPT_l_to_m}. In this case, the mean first passage time is a monotonic increasing function of the ensemble temperature. At high temperature, the mean fist passage time becomes very long. This implies that it is very difficult for the two black hole phase to escape to the stable wormhole phase. The reason that the kinetic time becomes longer at higher temperature is that the barrier height on the free energy landscape between the two black hole phase and the unstable wormhole phase increases monotonically with the ensemble temperature as shown in Figure \ref{barrier_height}. In Figure \ref{barrier_height_vs_MFPT_ltm}, we plot the correlation between the barrier height and the mean kinetic time $\langle t\rangle$ for the phase transition from the two black hole phase to the unstable wormhole phase. It is also shown that kinetic time is the monotonous function of the barrier height. Therefore, we can conclude that the free energy barrier is the dominate element that impacts the kinetic time of the transition process.

The behavior of the relative fluctuation as the function of the ensemble temperature is plotted on the right panel of Figure \ref{MFPT_l_to_m}. As discussed above, the relative fluctuation attains the maximum at the temperature where the mean first time is at its minimum. When the ensemble temperature approaches $T_1$, the kinetic time $\langle t \rangle$ attains its minimum, while the relative fluctuation attains its maximum. As the temperature increases, the relative fluctuation decreases from the maximum to the minimum. It should be pointed out that, because of the considerable numerical errors for the kinetic time $\langle t \rangle$ and $\langle t^2 \rangle$, it is difficult to compute the relative fluctuation precisely. However, the general trend is not affected by the numerical errors.

\section{Conclusion}\label{conclusion}

In the present work, we apply the stochastic dynamics method to study the phase transition in two coupled SYK models. For the two coupled SYK model at the large $N$ limit, according to the thermodynamics of the system, we define the off-shell free energy and quantify the corresponding free energy landscape. In the framework of the free energy landscape, we analyze the $\mu-T$ phase diagram and obtain the conclusion that the phase transition between the wormhole and two black hole in this model is analogous to the Van der Waals type, not exactly the Hawking-Page phase transition. Furthermore, we propose that the kinetics of the phase transition between the wormhole and two black hole can be investigated by the stochastic dynamics on the underlying free energy landscape. We assume the phase transition process between the wormhole and two black hole is a stochastic process under the thermal fluctuations and study the probabilistic Fokker-Planck equation which describes the kinetics of the phase transition process. By calculating the kinetics time and its fluctuation, we reveal the underlying thermodynamics and kinetics of the phase transition between the wormhole and two black hole. It is shown that the underlying reason of the kinetics behavior is determined by the barrier heights and the temperature on the free energy landscape.

\end{document}